\documentclass[12pt]{iopart}
\pdfoutput=1
\usepackage{cite}
\usepackage{graphicx}

\begin{document}

\title[Variational estimation of the drift for stochastic differential
equations]{Variational estimation of the drift for stochastic
  differential equations from the empirical density}
\author{Philipp Batz, Andreas Ruttor and Manfred Opper}
\address{Artificial Intelligence Group, Technische Universit\"{a}t
  Berlin, Marchstra{\ss}e 23, Berlin 10587, Germany}
\eads{\mailto{philipp.batz@tu-berlin.de},
  \mailto{andreas.ruttor@tu-berlin.de},
  \mailto{manfred.opper@tu-berlin.de}}

\begin{abstract}
  We present a method for the nonparametric estimation of the drift
  function of certain types of stochastic differential equations from
  the empirical density. It is based on a variational formulation of
  the Fokker-Planck equation. The minimization of an empirical
  estimate of the variational functional using kernel based
  regularization can be performed in closed form. We demonstrate the
  performance of the method on second order, Langevin-type equations
  and show how the method can be generalized to other noise models.
\end{abstract}

\pacs{02.50.Tt, 02.50.Ey, 05.10.Gg, 05.45.Tp, 45.10.Db}

\section{Introduction}

An important problem in modelling a random process by a stochastic
differential equation (SDE) is the fitting of the model to observed
data. An SDE is determined by its drift function and the diffusion.
For models of thermal equilibrium, where the diffusion is proportional
to the unit matrix and the drift is the gradient of a potential, a
rather simple and well known approach for estimating the drift from
data is available (see e.g.~\cite{Iacus:2008:SIS}): One can use the
fact that the potential is proportional to the logarithm of the
stationary density. Hence, from an estimator of the density, provided
e.g.~by a kernel density estimator (KDE) one can get an explicit
estimator for the drift. This estimator is based on the empirical
distribution of data alone which completely ignores the temporal
ordering of observations and the time lag between them. One only needs
an ergodic sample of the process. The KDE estimator is also
nonparametric i.e.~it does not assume a specific parametric functional
form of the drift. For non-equilibrium models such an explicit
expression of the density in terms of the drift is, in general, not
known. Also in higher dimensions the convergence of the KDE to the
true density with increasing data sample size maybe slow
\cite{Sriperumbudur:2014:DEI}.

Other parametric and nonparametric approaches to drift estimation have
to deal with the problem of low data sampling rates
\cite{Gottschal:l2008:DHD}. E.g.~a nonparametric method based on
Kramers-Moyal coefficients \cite{Honisch:2011:EKM} (conditional
moments) requires numerical solutions of the Kolmogorov backward
equation over a time interval given by the time lag. Bayesian
estimators using a Gaussian process prior over drift functions provide
an elegant solution to estimation \cite{Papaspiliopoulos:2012:NED}
when a complete path of dense observations is available. But in
general it requires the imputation of unobserved diffusion paths as
hidden random variables between neighboring observations. This can
lead to time consuming computations or requires further
approximations. \cite{Papaspiliopoulos:2012:NED} introduced a Monte
Carlo Gibbs sampler, which switches between sampling hidden paths of
the process and sampling drift functions. An alternative approach was
given in \cite{Ruttor:2013:ACP}, where the latent path was treated by
an expectation maximization approach with the hidden process
approximated by a linear stochastic differential equation. This seems
to work faster, but the quality of the linear approximation
deteriorates for larger time lags leading to an asymptotic bias in the
inference of the drift.

The goal of this paper is to construct classes of nontrivial SDE
models for which a computationally efficient nonparametric estimation
of the drift is possible using the empirical distribution alone. Our
method is based on a variational formulation of the stationary
Fokker-Planck equation which gives a unique solution to the drift
under certain conditions. These generalize the potential condition of
thermal equilibrium. The functional to be minimized is an expectation
over the stationary density. By replacing this density with the
empirical one, i.e.~with an unordered data sample and by minimizing
the empirical functional (e.g.~in a parametric family of potentials),
one can get an estimator of the drift. The method can be generalized
to a nonparametric estimate if the empirical functional is regularized
with a kernel based penalty term. Our approach is not based on an
explicit representation of the drift in terms of the density. Hence,
it does not use a direct estimator of the density such as the KDE.
Thus it is not expected to suffer from the bad convergence properties
of the KDE in higher dimensions.

The paper is organized as follows. The second chapter introduces the
variational formulation of the Fokker-Planck equation from which the
drift can be derived by minimization of a functional. The third
chapter shows how a regularized empirical approximation of the
functional leads to a nonparametric estimate. The fourth chapter
presents examples of this estimator for the class of Langevin
equations for which the extra conditions lead to only mild
restrictions. The fifth chapter explains how the method can be
extended to other types of noise, such as jump processes. We conclude
with a discussion and possible extensions of the method in chapter
six.

\section{A variational formulation for the Fokker-Planck equation}

We consider stochastic differential equations for the dynamics of a
$d$-dimensional diffusion process $Z_t\in R^d$ given by
\begin{equation}
  dZ_t = g(Z_t) dt  + \sigma(Z_t) dW_t.
  \label{eq:SDE}
\end{equation}
The drift function $g(\cdot)\in R^d$ represents the deterministic part
of the driving force and $W$ is a $k$-dimensional ($k\leq d$) vector
of independent Wiener processes acting as a white noise source. The
strength of the noise is determined by the state dependent $d\times k$
dimensional noise matrix $\sigma(Z)$.

Suppose that we are given the stationary density $p(x)$ of the
process. How can we determine the drift $g$ which corresponds to this
density? To give a partial answer to this question, we assume that
$\sigma(\cdot)$ is known and the drift splits into two parts $g(z) =
r(z) + f(z)$, where $r(z)$ is a known part and we try to compute $f$.
Of course in the multivariate case there is not enough information to
reconstruct $f$ uniquely. However we may search for a \emph{minimal}
solution which minimizes a quadratic functional
\begin{equation}
  \frac{1}{2} \int p(z)\; f(z) \cdot A^{-1} (z) f(z)\; dz
  \label{KLrate2}
\end{equation}
for a given positive definite matrix $A(z)$. Introducing a Lagrange
multiplier function $\psi(z)$ for the condition that the density $p$
fulfils the stationary Fokker-Planck equation with drift $g$, we can
derive the minimal $f$ from the variation of the Lagrange-functional
\begin{equation}
  \frac{1}{2} \int f(z) \cdot A^{-1} (z) f(z) \; dz - \int \psi(z)
  \left\{ \mathcal{L} p(z) - \nabla \cdot (f(z) p(z)) \right\} \; dz
  \label{Lagrange}
\end{equation}
where the Fokker-Planck operator $\mathcal{L}$ corresponding to the
known drift $r(z)$ is given by
\begin{equation}
  \mathcal{L} p(z) = -\nabla \cdot (r(z) p(z)) + \frac{1}{2} \tr
  \left[ \nabla \nabla^\top(D(z) p(z)) \right]
  \label{Fokker}
\end{equation}
with $D(z) \doteq \sigma(z) \sigma(z)^\top$. Variation of
(\ref{Lagrange}) with respect to $f$ yields $f(z) = A(z) \nabla
\psi(z)$. Inserting this solution back into (\ref{Lagrange}) shows
that the unknown 'potential' $\psi$ can be derived from the
minimization of the functional
\begin{equation}
  \varepsilon[\psi] = \int \left\{ \frac{1}{2} \nabla \psi(z) \cdot
    A(z) \; \nabla \psi(z) + \mathcal{L}^* \psi(z) \right\} p(z) dz
  \label{epsicost}
\end{equation}
where $\mathcal{L}^*$ is the adjoint operator of $\mathcal{L}$,
(\ref{Fokker}) which fulfils
\begin{equation}
\int \psi(z) \mathcal{L} p(z) dz = \int
p(z) \mathcal{L}^*\psi(z) dz
\end{equation}
and is given by
\begin{equation}
  \mathcal{L}^* \psi(z) = r(z) \cdot \nabla \psi(z) + \frac{1}{2} \tr
  \left[ D(z) \nabla \nabla^\top \psi(z) \right]
  \label{adjoint}
\end{equation}
In fact, a direct minimization of (\ref{epsicost}) with respect to
$\psi$ yields
\begin{equation}
  \mathcal{L}[\psi] p(z) \doteq \mathcal{L} p(z) - \nabla \cdot \left(
    A(z) \nabla \psi(z) p(z) \right) = 0.
  \label{eq:FPE1}
\end{equation}
which is the stationary Fokker-Planck equation corresponding to the
density $p(z)$ and the drift $g(z) = r(z) + A (z) \nabla \psi$. Hence,
if the drift is actually of this form, then the minimization of
(\ref{epsicost}) will give us the desired unique result. For the
special case $D= A = I$ and $r=0$ the functional (\ref{epsicost}) was
introduced in the field of machine learning as a \emph{score-function}
for estimating $\ln p(x)$ up to a normalization constant
\cite{Hyvarinen:2005:ENN}. This case corresponds to an SDE for thermal
equilibrium where the drift $f(z) = \nabla \psi(z)$ is the gradient of
a potential $\psi$ and the stationary density fulfils $p(z) \propto
e^{2 \psi(z)}$.

The matrix $A$ introduces an extra degree of freedom which could be
chosen using prior knowledge of the SDE model. Of special interest are
models with $A = D$. As we show in \ref{sec:posterior}, for such
models we have asymptotically $\varepsilon[\psi] \simeq
\frac{1}{T}\varepsilon_{ML}[\psi] $, where $\varepsilon_{ML}$ is the
negative log-likelihood, when the process $Z_t$ was sampled
continuously in time over a large time $T$. Hence if observations are
dense in time a minimization of $\varepsilon[\psi]$ using the
empirical distribution should become asymptotically equivalent to
maximum likelihood estimation. The discussion in \ref{sec:posterior}
also gives another interpretation of the cost function
(\ref{epsicost}) for $A=D$. The drift given by the minima of
(\ref{epsicost}) leads to the process with path measure that is
closest in relative entropy (Kullback-Leibler divergence) rate
\cite{Archambeau:2007:VIF, Chernyak:2013:SOC} to the path measure of
the process with drift $r(z)$, when $p(z)$ is given.

\section{Minimizing the empirical functional}

Our goal is to estimate $\psi$ from data by replacing the average over
the stationary density $p(z)$ in the functional (\ref{epsicost}) by
the empirical distribution
\begin{equation}
  \hat{p}(z) = \frac{1}{n} \sum_{i=1}^n \delta(z - z_i)
  \label{emp_density}
\end{equation}
where $z_1, \ldots, z_n$ is a random, ergodic sample drawn from this
density. An obvious possibility to construct estimators is to work
with a parametric representation
\begin{equation}
  \psi_w(z) = \sum_{k=1}^K w_k \phi_k(x)
  \label{funcrep}
\end{equation}
where the $\phi_k$ are set of given 'basis' functions. The weights
$w_k$ could be determined by minimization of the empirical version of
the functional (\ref{epsicost})
\begin{equation}
  \varepsilon_{emp}[\psi_w] = \sum_{i=1}^n \left\{ \frac{1}{2} \nabla
    \psi_w(z_i) \cdot A(z_i) \; \nabla \psi_w(z_i) + \mathcal{L}^*
    \psi_w(z_i) \right\}
  \label{statlim}
\end{equation}
which is a quadratic form in the $w_k$ and can thus be performed in
closed form. We are however interested in the case where a
representation in terms of a {\em finite} set of basis functions is
not rich enough to represent $\psi$. Thus we will resort to a more
general, nonparametric representation allowing for an infinite set of
functions $\phi_k$. Since one has only a finite number of data $z_i$
for estimation, the estimator needs to be regularized by introducing
an extra penalty term. This will be chosen as a quadratic form $
\frac{1}{2}\sum_k w_k^2 / \lambda_k$, where the $\lambda_k$ are
hyper-parameters. This penalty can also be viewed from a
pseudo-Bayesian perspective where $\exp\{-
C\varepsilon_{emp}[\psi_w]\}$ is interpreted as a likelihood and
$\exp\{-\frac{1}{2} \sum_k w_k^2 / \lambda_k\}$ as a Gaussian prior
distribution over parameters $w_k$. $C$ can be chosen to give
different weight to the data and to the penalty. In this
interpretation, (\ref{funcrep}) could be understood as a Gaussian
process model \cite{Rasmussen:2006:GPM} for the function $\psi$. As
shown in \ref{sec:posterior}, the likelihood interpretation becomes
correct asymptotically for densely sampled observations if $A = D$ for
which we would set $C\approx T/n$ being the time between observations.

Motivated by the Gaussian process point of view we will introduce the
\emph{kernel trick} into our formalism avoiding an explicit
specification of $\phi_k$ and $\lambda_k$ and assume instead that
these are defined implicitly as orthonormal eigenfunctions and
eigenvalues of a positive definite kernel function $K(z,z')$ via
\begin{displaymath}
  K(z,z') \doteq \sum_k \lambda_k \phi_k(z) \phi_k(z')
\end{displaymath}
This can be viewed as the covariance kernel of a Gaussian process
prior distribution for functions $\psi(z)$ \cite{Rasmussen:2006:GPM}.
Kernels can be adapted to the prior knowledge which is available about
the function $\psi$. This might include a known periodicity of the
function, the length scale of its typical variation, or the fact that
$\psi$ is a polynomial of a given order. \ref{sec:kernel} gives a
short summary of the kernels used in our experiments.

In the kernel approach the regularized functional can be written as
\begin{eqnarray}
  &C& \sum_{i=1}^n \left\{ \frac{1}{2} \nabla \psi(z_i) \cdot A(z_i) \;
    \nabla \psi(z_i) + \mathcal{L}^* \psi(z_i) \right\}\nonumber\\
    && + \frac{1}{2}
  \int \int \psi(z) K^{-1}(z,z') \psi(z') \; dz dz',
  \label{regfunc}
\end{eqnarray}
where $K^{-1}(z,z')$ is the inverse of the kernel operator. One can
also show that the penalty term on the right hand side equals the
so-called {\em reproducing kernel Hilbert space} (RKHS) norm of $\psi$
defined by the kernel $K$. Using this formalism a nonparametric
extension of the score function approach for estimating $\ln p(x)$ was
introduced in \cite{Sriperumbudur:2014:DEI}. Our discussion shows,
that there are two ways for computing the estimator of $\psi(z)$
explicitly, both leading to the same result. The first one is based on
setting the variational derivative of (\ref{regfunc}) equal to zero
and the second uses the formalism of Gaussian process regression
\cite{Rasmussen:2006:GPM}. We will next give a derivation of this
result using the first method. Performing the variation with respect
to $\psi$ yields
\begin{displaymath}
  C n \mathcal{L}[\psi] \hat{p}(z) + \int K^{-1}(z,z') \psi(z') \; dz'
  = 0,
\end{displaymath}
where $ \mathcal{L}[\psi]$ was defined in (\ref{eq:FPE1}). Multiplying
both sides of this equation with the operator $K$ we get
\begin{equation}
  \psi(z) + C \sum_{j=1}^n \mathcal{L}^*_{z'}[\psi] \;
  K(z,z')_{z'=z_j} = 0
  \label{psiexplic}
\end{equation}
where the adjoint operator acts on functions $h$ as
\begin{equation}
  \mathcal{L}^*_{z'}[\psi] h(z') = (r(z') + A(z') \nabla \psi(z'))
  \nabla h(z') + \frac{1}{2} \tr \left[ D(z') \nabla \nabla^\top h(z')
  \right]
  \label{adjoint_psi}
\end{equation}
We can understand (\ref{psiexplic}) as a regularized version of the
equation
\begin{equation}
\int p(z') \mathcal{L}^*_{z'}[\psi] h(z') dz' = \int h(z')
\mathcal{L}_{z'}[\psi] p(z') dz' = 0
\end{equation}
applied to the family of kernel functions $h_z(z') = K(z,z')$ when the
stationary density $p$ is replaced by its empirical approximation
$\hat{p}$ (\ref{emp_density}).

Equations (\ref{psiexplic}) and (\ref{adjoint_psi}) show that if
$\nabla \psi(z)$ is known at all sample points $z=z_i$, we can
evaluate the second term and get then the function $\psi(z)$ for all
$z$. The gradient of $\psi$ at the data points is computed by taking
the gradient of (\ref{psiexplic}) and setting $z=z_i$. This yields the
set of linear equations
\begin{equation}
  \nabla \psi(z_i) + C \sum_{j=1}^n \mathcal{L}^*_{z'}[\psi]
  \nabla_{z} K(z,z')_{z= z_i, z'=z_j} = 0
\end{equation}
for the $d\times n$ unknowns $\nabla \psi(z_i)$ which can be plugged
into (\ref{psiexplic}) to obtain the explicit result for the
estimator.

\section{Application: Langevin dynamics}

To show that the condition $f(z) = A(z)\nabla \psi(z)$ includes
classes of nontrivial non-equilibrium models, we will specialize to
second order (Langevin-type) SDE which appear naturally when systems
of classical mechanics are driven by deterministic and random forces.
The time evolution is described in terms of (generalized) coordinates
and velocities $X, V\in R^d$ as
\begin{equation}
  dX_t = V_t dt, \qquad
  dV_t = g_v(X_t, V_t) dt + \sigma_v (X_t, V_t) dW_t.
  \label{Langevin}
\end{equation}
The noise acts only on the acceleration and the drift for this model
is of the form $(g_x, g_v)$ where $g_x = v$ is known. Hence, we may
choose the matrix elements of $A$ in (\ref{epsicost}) to be zero
except for the sub-matrix $A_{vv}$ which we will set to the unit
matrix $A_{vv} = I$ for simplicity. Thus the reduced drift vector for
the velocity is assumed to be of the form $g_v(x,v) = r_v(x,y) +
\nabla_v \psi(x,v)$ and the functional (\ref{epsicost}) becomes
\begin{equation}
  \varepsilon[\psi] = \int p(x,v) \left\{ \mathcal{L}^* \psi(x,v) +
    \frac{1}{2} (\nabla_v \psi(x,v))^2 \right\} dx \, dv
  \label{costfu2D}
\end{equation}
where
\begin{equation}
  \mathcal{L}^* \psi(x,v) = \left( v \cdot \nabla_x + r_v(x,v) \cdot
    \nabla_v + \frac{1}{2} \tr(D_v(x,v) \nabla_v^\top \nabla_v)
  \right) \psi (x,v),
  \label{adjoint_Langevin}
\end{equation}
with $D_v = \sigma_v \sigma_v^\top$. The \emph{integrability}
condition on the unknown part of the drift $f_v(x,v) = \nabla_v
\phi(x,v)$ restricts the velocity dependency rather than its
coordinate dependency. We will specialize on dynamical systems with a
position dependent external force $f(x)$ and a friction term which is
linear in the velocity. This is given by
\begin{equation}
  f_v(x,v) = f(x) - \Lambda v = \nabla_v \left\{v\cdot f(x) -
    \frac{1}{2} v \cdot \Lambda v \right\},
  \label{drift_Langevin}
\end{equation}
with a positive diagonal matrix $\Lambda$ and an \emph{arbitrary}
(e.g.~non-conservative) vector field $f(x)$. Since the velocity
dependency is parametric, we use a kernel which is a product of a
first order polynomial kernel in $v$ and an RBF kernel (\ref{eq:rbf})
in $x$ and set $r=0$ to estimate $\Lambda$ and $f(\cdot)$ from $n$
pairs of observations $(x_i, v_i)$.

We illustrate this method for the case of a bistable system with two
locally stable equilibria (double well model) which corresponds to the
drift $f(x) = -4(x - x^3)$. We simulated the process with diffusion
$\sigma = 1.3$ and friction constant $\lambda = 1.1$ and generated a
data set of size $n = 3000$ observations with time lag $\tau = 0.5$.
We found that the constant $C$ in (\ref{psiexplic}) did not have a
strong influence on the accuracy of the estimator and have used $C=1$
throughout the experiments. In order to optimize the length scale
hyperparameter of the RBF kernel, we randomly divided the observation
into two subsets of equal size and used a conjugate gradient
optimization to minimize the cost function (\ref{costfu2D})
approximated by the hold out data.

\begin{figure}
  \centering
  \includegraphics[width=0.4\textwidth]{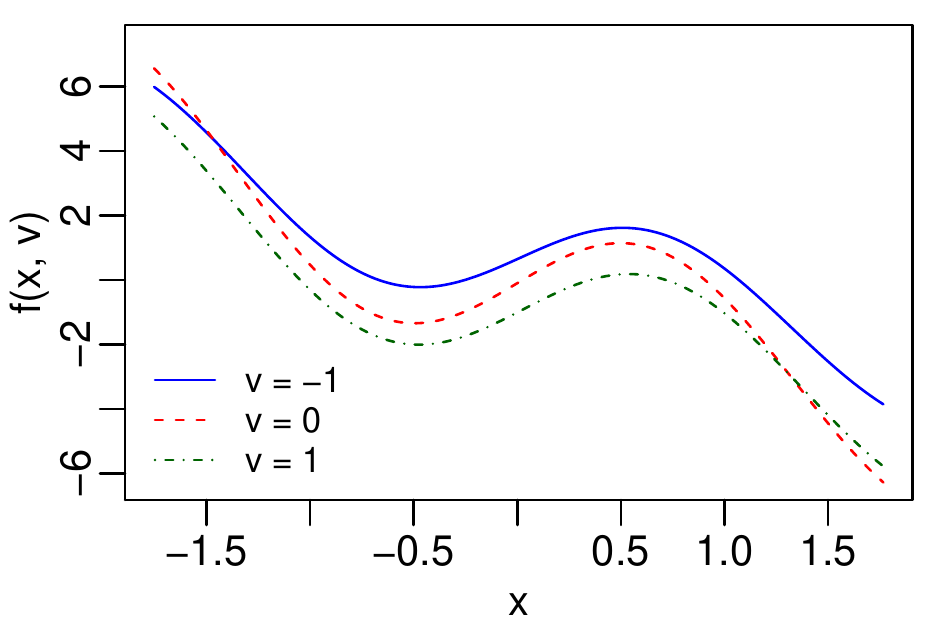} % DWDriftxgivenv
  \includegraphics[width=0.4\textwidth]{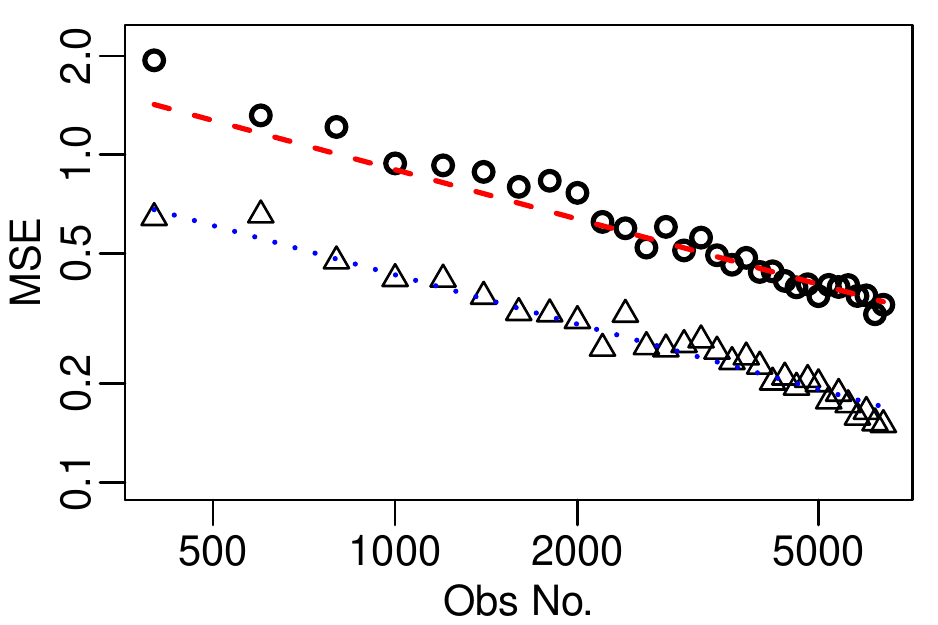} % DWLearningCurve
  \caption{Left: Estimates of the double well drift function $f(x, v)$
    as a function of the coordinate $x$ with fixed velocities $v$. The
    RBF kernel uses the length scale $l_{opt} = 1.29$. Right: MSE of
    the estimator on hold out data (prediction error, triangles) and
    on the observations (training error, circles) for the double well
    model as a function of number $n$ of observations. The lines in
    this log-log plot were fitted using the model $\textit{MSE}
    \propto n^{-1/2}$. Each point represents the average over $m = 5$
    different data sets.}
  \label{fig:Doublewell}
\end{figure}

The left of figure \ref{fig:Doublewell} shows the estimate $f_v(x,v)$
as a function of the state $x$ with the velocity $v$ fixed to three
different values. We also studied a measure of convergence of our
estimator $f_v$ to the true drift for the double well model. For that
purpose we have computed the mean squared error (MSE) for different
number $n$ of observations, both as training error at the observed
data points and as prediction error estimate on a hold out data set.
We have used a Gaussian process with RBF kernel and length scale $l =
1.25$. The data sets have been generated with diffusion constant
$\sigma = 1$, friction constant $\lambda = 1.1$, and time lag $\tau =
4$. The resulting learning curves are shown in the right of figure
\ref{fig:Doublewell}. Comparing the asymptotic power law fits for both
MSEs shows that, while naturally the prediction error has a higher
baseline error rate than the training error, both are consistent with
a decay of the form $\approx n^{-1/2}$.

\begin{figure}
  \centering
  \includegraphics[width=0.5\textwidth]{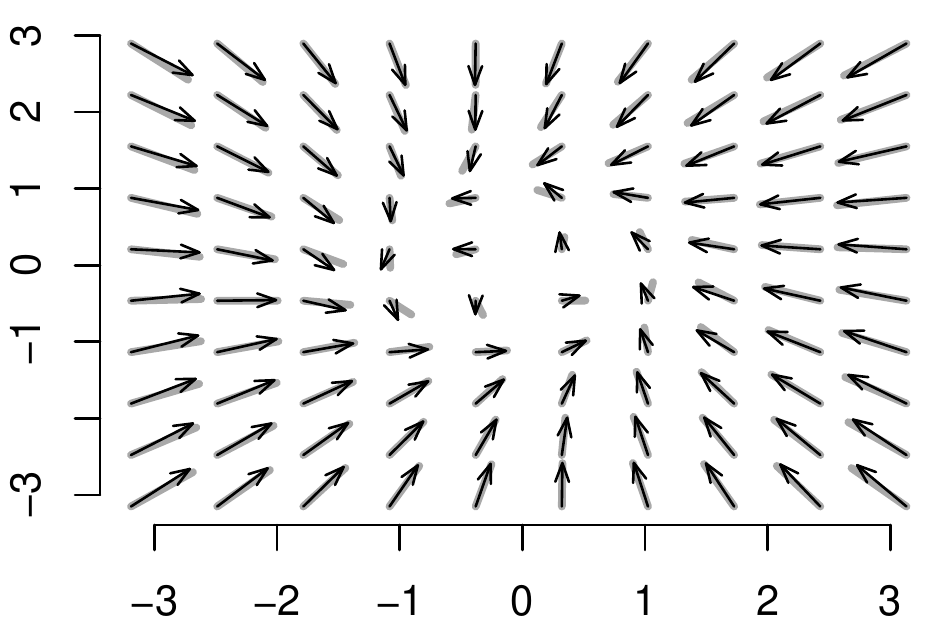} % 2DDWEstTrueVF
  \caption{The figure shows the vector fields of both the true drift
    function given as grey lines and the estimated drift function
    given as arrows. In order to enhance visibility, we scaled the
    length of the vectors logarithmically.}
  \label{fig:2DVF}
\end{figure}

So far we have assumed that the diffusion $D_v$ is known. If on the
other hand, the friction parameter $\Lambda$ is known, i.e.~$r_v = -
\Lambda v$ then $\mathcal{L}^* \psi(x,v) = \mathcal{L}^* (v\cdot
f(x))$ is independent of the diffusion term $D_v(x,v)$ and we can
estimate $f(x)$ without knowing the diffusion. We demonstrate this
estimate on a two dimensional Langevin model with a nonconservative
drift with components $f^{(1)}(x) = x^{(1)}(1 - (x^{(1)})^2 -
(x^{(2)})^2) - x^{(2)}$ and $f^{(2)}(x) = x^{(2)}(1 - (x^{(1)})^2 -
(x^{(2)})^2) - x^{(1)}$, and friction constants $\lambda^{(1,2)} =
0.8$. The $n=2000$ four-dimensional data (position and velocity)
observations were generated with constant diffusion constants $D_v = 9
I$ and time lag $\tau = 0.5$. The drift vector was estimated by
penalizing each component of $f(x)$ independently using a polynomial
kernel (\ref{eq:poly}) of order $p=4$, assuming that the true drift is
at most a polynomial of order 4. The results are shown in figure
\ref{fig:2DVF}.

\begin{figure}
  \centering
  \includegraphics[width=0.4\textwidth]{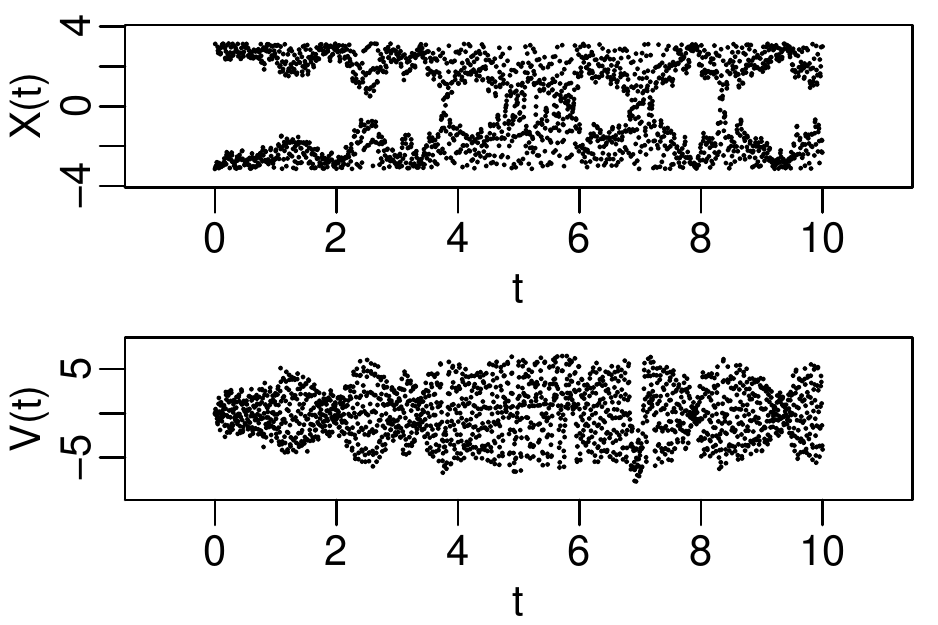} % CartPoleTrajectory
  \includegraphics[width=0.4\textwidth]{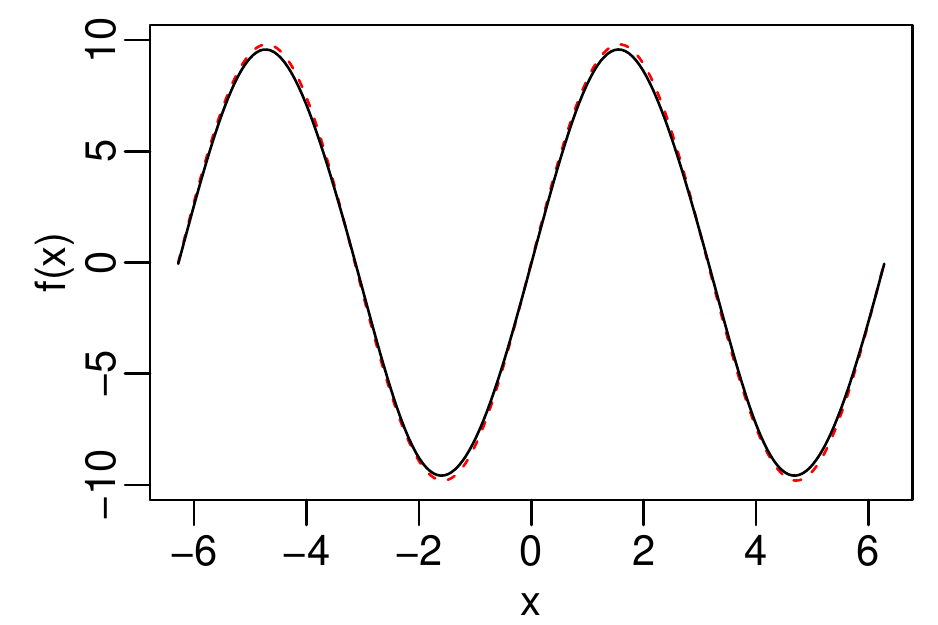} % CartPoleDriftFullGP
  \caption{Left: Snippet of the full sample path of the Cart and pole
    model. Right: Estimated drift function for Cart and pole, where
    the red dashed line denotes the true drift function and the black
    line the estimate.}
 \label{fig:CPDrift}
\end{figure}

To show that this approach also works for state dependent diffusion
(multiplicative noise), we consider a model of a pendulum on which
gravitation and friction act as drift terms given by $f(x) = a \sin x$
and $-\lambda v$ in (\ref{drift_Langevin}). Here $x$ is the angle
relative to the upward position and $v$ the angular velocity. The
pendulum is mounted on a cart that is accelerated \footnote{This toy
  model is known as \emph{cart and pole}, frequently used to test
  control methods for stabilizing the pendulum in the upright position
  \cite{Deisenroth:2009:GPD}.} in the horizontal direction by a white
noise force. This leads to an additional stochastic angular
acceleration with diffusion $D_v = (\sigma\cos(x))^2$. For the
simulation, we used a data set of $n = 2000$ observations with time
lag $\tau = 0.25$, diffusion $\sigma = 1$, $a = 9.81$ and $\lambda =
0.05$. As kernel function we chose a periodic kernel
(\ref{eq:periodic}) with hyperparameter $l_\mathrm{Per} = 1.2$. One
can clearly see from the left of figure \ref{fig:CPDrift} that most of
the time the pendulum stays in the downward position and only
occasionally crosses the upright position (corresponding to $x = 0$).
Nevertheless, the right panel of figure \ref{fig:CPDrift} shows that
regularization with the periodic kernel leads to an excellent
estimation of the drift for all values of $x$.

One might wonder if the estimation of the drift could have also be
achieved by a kernel density approach. While for a general model
(\ref{eq:SDE}) with drift $f(z) = r(z) + A(z) \nabla \psi(z)$ there
does not seem to be a way of expressing $f(z)$ in terms of $p(z)$ in
closed form, a somewhat complicated expression can be given for
Langevin equations with drifts of the form (\ref{drift_Langevin}).
Multiplying the Fokker-Planck equation for the process
(\ref{Langevin}) with a component $v^{(i)}$ of the velocity vector and
integrating over $v$, one obtains the following explicit
representation for the drift
\begin{equation}
  f^{(i)}(x) = \sum_{j=1}^d \frac{\partial E[v^{(i)} v^{(j)}
    |x]}{\partial x^{(j)}} + \sum_{j=1}^d E[v^{(i)} v^{(j)} |x]
  \frac{\partial \ln p(x)}{\partial x^{(j)}} - E[r^{(i)} | x],
  \label{explic}
\end{equation}
where components of vectors are denoted by superscripts and $E[\cdot |
x]$ denote conditional expectations. This shows that, in general, one
would need not only a KDE for estimating $p(x)$ but also a
nonparametric regression method for estimating the conditional
expectations as a function of $x$. Of course, for the equilibrium case
where $f(x) = \nabla \phi$ and $D_v \propto \Sigma$, $r = - \Lambda v$
where $\Lambda$ and $\Sigma$ are diagonal matrices which satisfy
$2\lambda_i /\sigma_i^2 = \beta$, which is the inverse temperature,
(\ref{explic}) simplifies because one has $E[v^{(i)} v^{(j)} |x] =
\frac{1}{2\beta} \delta_{ij}$ and $E[r^{(i)} | x] = 0$. For this case,
the velocity samples $v_i$ are not needed.

\section{Generalization to other noise processes}

For applications where noise is used as a part of an external control
signal acting on a dynamical system, the assumption of white noise is
not realistic, because its non-decaying high frequency components are,
in practice, filtered out in the control circuit. Hence, we would like
to include other processes, e.g.~colored noise or a noise source with
a finite state space. To adapt our method to this situation, we
replace the white noise $\sigma_v(x,v) dW$ in (\ref{Langevin}) by $u_t
dt$ where $u_t$ is a Markov process, which is included in the state
variable $z = (x,v,u)$ and observed at the same time times as $X$ and
$V$. Our formalism does not change when $u$ is a diffusion process
itself, because the entire system can be described by a Fokker-Planck
equation for the density $p(x,v,u)$. But it is also possible to
include other Markov processes such as jump processes as noise
sources. We just have to replace the Fokker-Planck equation by the
appropriate Master equation in the definition (\ref{Fokker}). We will
illustrate this idea for $u_t$ being a random telegraph process
\cite{Gardiner:1996:HSM} that switches with equal rates $\gamma$
between $u = \pm 1$. We study a one-dimensional system $x,v,u \in R$
with drift given by $g_v(x,v) = - \lambda v + f(x)$ with a known
friction constant $\lambda$. The \emph{Master equation}
\cite{Gardiner:1996:HSM} for the stationary density is given by
\begin{eqnarray}
  &-&\partial_x v p(x,v,u) - \partial_v \left[ (f(x) - \lambda v + u)
    p(x,v,u) \right] \nonumber \\
    &+& \gamma \left(p(x,v,-u) - p(x,v,u) \right) = 0
  \label{Master:jump}
\end{eqnarray}
The drift $f$ can be estimated by the minimization of the functional
(\ref{costfu2D}), when we use the adjoint operator given by
\begin{displaymath}
  \mathcal{L}^*\psi(x,v,u) = \left\{ v \partial_x + (u - \lambda
    v) \partial_v \right\} \psi(x,v,u) + \gamma \left( \psi(x,v,-1) -
    \psi(x,v,1) \right)
\end{displaymath}
The  	parameterization $\psi(x,v) = v f(x)$ leads to the functional
\begin{equation}
  \varepsilon[f] = \frac{1}{2}\sum_{u = \pm 1} \int p(x,v,u) \left\{
    f^2(x) + 2 f'(x) v^2 + 2 f(x) (u - \lambda v) \right\} dx dv
\end{equation}
to be minimized with respect to $f$. Experiments (not included here)
for the \emph{cart and pole} model show that this method achieves
similar performance as the one shown in figure \ref{fig:CPDrift}.

\section{Discussion and Outlook}

We have presented a method for a nonparametric estimation of the drift
of certain types of stochastic differential equations from the
empirical density alone. The method is not related to kernel density
estimation and can be applied to cases where the use of a KDE would
not be simple or impossible. The method should be of interest for
situations where external noise is used to explore the state space of
a mechanical system in order to \emph{learn} the deterministic part of
the forces which can be used to later control the system. On the other
hand, one might use our variational approach for solving a specific
type of stochastic control problem \cite{Chernyak:2013:SOC}: We would
be able to compute a state dependent control $f(z)$ which has to be
added to the known drift $r(z)$ of a system such that a new desired
stationary density $p(z)$ will be reached.

In future work we will explore different possibilities to increase the
applicability of our methods. We will investigate carefully the role
of the matrix valued model parameter $A$ in (\ref{Fokker}) but
also try to generalize the functional (\ref{epsicost}) by including
other types of operators. E.g.~a second derivative of a convex
potential $\psi$ could be used to estimate the diffusion for known
drift. It will also be interesting to analyze the theoretical
properties of the estimator, especially the asymptotic convergence
rate towards the \emph{true} drift function.

Another question is how to lift the restriction that all coordinates
of the random state vector $z$ need to be observed jointly. For the
Langevin type equations with a drift of the form
(\ref{drift_Langevin}) it would be interesting to see if the method
could be generalized to estimating a drift $f(x)$ based on coordinate
observations $x_i$ alone. For potentials of the type $v\cdot f(x)$ one
can integrate over the velocities in (\ref{costfu2D}) to obtain a
functional which depends on $f(x)$, $p(x)$ and conditional moments of
the velocities (see (\ref{explic})). For an equilibrium problem, these
conditional moments are constant and known and the velocity
observations are not needed. For the general case one could use the
temporal order of coordinate observations to obtain a preliminary
approximation to the unobserved velocities. An initial estimate of the
drift $f(x)$ could then be derived by a minimization of the functional
(\ref{costfu2D}). This estimate could be used to create new velocity
samples and estimates for conditional velocity moments by performing
forward sampling of the SDE (\ref{Langevin}) and the method could be
iterated. Preliminary experiments using this iterative procedure are
promising but we do not yet have conditions on the convergence of such
a procedure.

\ack

This work was supported by the European Community's Seventh Framework
Pro\-gramme (FP7, 2007-2013) under the grant agreement 270327
(CompLACS).

\appendix

\section{Likelihood for dense observations}
\label{sec:posterior}

We will derive the likelihood function for the drift $g$ of an SDE
assuming that we have access to a {\em dense} path $Z_{0:T}$ of
observations in a time window from $t=0$ to $t=T$. Discretizing time
into small intervals of length $\Delta t$ and using the fact that for
$\Delta t\to 0$, the transition density of (\ref{eq:SDE}) becomes
Gaussian, we find that the part of the negative log-likelihood (NLL)
which depends on the drift function $g$ can be approximated by
\begin{equation}
  -\ln p(Z_{0:T} | g) \simeq \frac{1}{2} \sum_t  \left\{ ||g (Z_t)||^2
    \Delta t - 2 \langle g(Z_t), (Z_{t + \Delta t} - Z_t) \rangle
  \right\} + \mbox{const}
  \label{eq:likelihood}
\end{equation}
where we have introduced the inner product $\langle u, v \rangle
\doteq u^\top D^{-1} v$ and the corresponding squared norm $||u||^2
\doteq u^\top D^{-1} u$. For $\Delta t \to 0$, the second sum becomes
a Ito stochastic integral \cite{Gardiner:1996:HSM}. If the drift can
be written as $g = r + D \nabla \psi$ where $r$ is a known function
and we are only interested in estimating $\psi$, we can transform the
Ito integral into an ordinary time integral using Ito's formula. The
part of the NLL which contains $\psi$ becomes
\begin{eqnarray}
  && \varepsilon_{ML}[\psi] = \frac{1}{2} \int_0^T
  \left\{ \nabla \psi \cdot D \; \nabla \psi \, dt + 2 r\cdot \nabla
    \psi \, dt - 2 \nabla \psi \cdot dZ_t \right\}
  \label{eq:full_likelihood} \\
  &=& \frac{1}{2} \int_0^T \left\{ \nabla \psi \cdot D \; \nabla \psi
    + 2 r \cdot \nabla \psi + \tr(D \nabla \nabla^\top \psi) \right\}
  dt - \psi(Z_T) + \psi(Z_0)
  \nonumber
\end{eqnarray}
We will now assume that for large $T$, the process becomes stationary
with density $p(z)$. We can then replace the time integral by an
integral over $p$. For a mathematical rigorous treatment see e.g
\cite{Papaspiliopoulos:2012:NED}. Neglecting the contribution from the
boundary terms in (\ref{eq:full_likelihood}) for large $T$ we arrive
at
\begin{equation}
 \varepsilon_{ML}[\psi] \simeq  \frac{T}{2} \int \left\{ \nabla \psi
   \cdot D \; \nabla \psi + 2 r \cdot \nabla \psi + \tr(D \nabla
   \nabla^\top \psi) \right\} p(z) dz.
\end{equation}
Using the Gaussian form of the transition density for short times
$\Delta t$ it can also be shown that the relative entropy or
Kullback-Leibler (KL) divergence between the path probabilities for
two diffusion processes but different drifts $g(z)$ and $r(z)$, where
$g(z) = r(z) + f(z)$ is given by
\begin{displaymath}
  D(p(Z_{0:T} | g) || p(Z_{0:T} | r) = \int_0^T dt \int p_t(z) f(z)
  \cdot D^{-1} (z) f(z) dz
\end{displaymath}
assuming they have the same diffusion term $D(z)$ and the same
non-random initial state (see e.g.~\cite{Archambeau:2007:VIF}). Here
$p_t(z)$ is the marginal density of the process with drift $g$ at time
$t$. Hence, assuming that the process becomes stationary with density
$p(z)$, we get for the relative entropy rate
\begin{equation}
  \lim_{T\to\infty} \frac{1}{T} D(p(Z_{0:T} | g) || p(Z_{0:T} | r) =
  \int p(z) f(z) \cdot D^{-1}(z) f(z) dz.
\end{equation}
A comparison with (\ref{Lagrange}) shows that for $A =
D$ the minimization of (\ref{epsicost}) leads to a process with given
stationary density that is closest to the process with drift $r(z)$ in
relative entropy. Hence, this may be understood as a generalized
\emph{maximum entropy} (minimum relative entropy) solution where the
stationary density is given as a constraint.

\section{Kernel functions}
\label{sec:kernel}

For the experiments we have used the following kernels:
\begin{itemize}
\item The radial basis function (RBF) kernel
  \begin{equation}
    K_\mathrm{RBF}(x, y) = \exp \left( -\frac{(x - y)^\top (x - y)}{2
        l_\mathrm{RBF}^2} \right),
    \label{eq:rbf}
  \end{equation}
  This has the length scale $l_\mathrm{RBF}$ as a hyper parameter and
  is used for estimating smooth functions.
\item The (one-dimensional) periodic kernel
  \begin{equation}
    K_\mathrm{Per}(x, y) = \exp \left( -\frac{2 \sin \left(\frac{x -
            y}{2} \right)^ 2}{ l_\mathrm{Per}^2} \right)
    \label{eq:periodic}
  \end{equation}
  is used for estimating smooth periodic functions.
\item The polynomial kernel of degree $p$
  \begin{equation}
    K_\mathrm{Poly}(x, y) = \left(1 + x^\top y \right)^p
    \label{eq:poly}
  \end{equation}
  is used for estimating functions which are polynomials with degrees
  at most $p$.
\end{itemize}

\section*{References}

\bibliographystyle{iopart-num}
\bibliography{nonpardrift}

\end{document}